\documentclass[12pt]{iopart}

\usepackage{color} 
\usepackage{subfig}  
\usepackage{graphicx} 
\usepackage{bm}
 
\usepackage[colorlinks=true,linkcolor=green,citecolor=blue]{hyperref} 

\begin{document}
\title[Effects of Helium massive gas injection level on disruption mitigation on EAST]{Effects of Helium massive gas injection level on disruption mitigation on EAST}
\author{Abdullah ZAFAR$^1$, Ping ZHU$^{2,3\ast}$, Ahmad ALI$^4$, Shiyong ZENG$^5$, and Haolong LI$^5$}

\address{\small $^1$ CAS Key Laboratory of Geospace Environment and Department of Engineering and Applied Physics, University of Science and Technology of China, Hefei 230026, China}
\address{$^2$ International Joint Research Laboratory of Magnetic Confinement Fusion and Plasma Physics, State Key Laboratory of Advanced Electromagnetic Engineering and Technology, School of Electrical and Electronic Engineering, Huazhong University of Science and Technology, Wuhan, Hubei 430074, China}
\address{$^3$ Department of Engineering Physics, University of Wisconsin-Madison, Madison, Wisconsin 53706, USA}
\address{$^4$ Pakistan Tokamak Plasma Research Institute, Islamabad 3329, Pakistan}
\address {$^5$ CAS Key Laboratory of Geospace Environment and School of Nuclear Science and Technology, University of Science and Technology of China, Hefei 230026, China}
\ead{zhup@hust.edu.cn}

\begin{abstract}\\
In this study, NIMROD simulations are performed to investigate the effects of massive Helium gas injection level on the induced disruption on EAST tokamak. It is demonstrated in simulations that two different scenarios of plasma cooling (complete cooling and partial cooling) take place for different amounts of injected impurities. For the impurity injection above a critical level, a single MHD activity is able to induce a complete core temperature collapse.  For impurity injection below the critical level, a series of multiple minor disruptions occur before the complete thermal quench (TQ).

\vspace{2pc}
\end{abstract}

\section{Introduction}

The most encouraging methods of disruption mitigation include the injection of impurities into the tokamak before an anticipated disruption. An efficient disruption mitigation system can spread the radiated power uniformly over the first wall, immensely reducing the heat load to the PFCs. Preferably, the time scales of both thermal and current quench will be sufficiently fast to dissipate the plasma energy prior to the development of disruptive instabilities, but slow enough to prevent the generation of  runaway electrons or substantial electromagnetic forces on the system~\cite{hender2007, lehnen2015}. Several mitigation techniques such as MGI (massive gas injections)~\cite{taylor1999, bakhtiari2005, granetz2006, pautasso2009, reux2010, lehnen2011}, shell-pellet injection~\cite{hollmann2009, hollmann2019}, and SPI (shatter-pellet injection)~\cite{commaux2010, commaux2016} have been carried out on DIII-D~\cite{hollmann2011}, Alcator C-MOD~\cite{bakhtiari2011}, ASDEX Upgrade~\cite{pautasso2011}, JET~\cite{lehnen2013, de2012}, EAST~\cite{duan2015, chen2018}, and J-TEXT~\cite{huang2018, ding2018} among other experiments.


To understand these experiments, various simulation schemes have been introduced for different timescales and phenomena involved in the disruption mitigation process. For example, 2D transport codes e.g. TOKES~\cite{pestchanyi2015}, and TokSys~\cite{hollmann2012} have been used to model the time scale of pre-TQ phase, whereas other 1.5D axisymmetric integrated modeling codes such as ASTRA~\cite{putvinski2010} and DINA~\cite{sugihara2012} allow the simulation of complete disruption phase in a moderate computational time span. The non-linear 3D MHD codes such as NIMROD~\cite{sovinec2004}, JOREK~\cite{huysmans2007, czarny2008} and M3D-C1~\cite{jardin2012} can simulate the time evolution of disruption mitigation when coupled with adequate impurity radiation and transport models. In particular, NIMROD, together with the coronal impurity dynamics model KPRAD (Killer Pellet RADiation), has been widely used for the MGI simulations~\cite{izzo2008, izzo2013, izzo2015}.


Recently, effects of different injection levels on TQ onset and radiated energy losses have been studied in disruption mitigation experiments with high pressure low-Z noble gas injection on EAST~\cite{duan2015, chen2018}, where the injected gas quantity primarily affects the duration of pre-TQ stage and the radiated energy loss. In this work, we perform NIMROD simulations to study the effect of impurity injection level on the MGI process based on an EAST discharge. We find that different levels of MGI trigger toroidal modes with different amplitudes and can indeed significantly modify the pre-TQ duration and radiated energy. In particular, we have identified a critical injection level which separates two distinctively different regimes for the MGI disruption mitigation processes.


This paper is organized as follows. The simulation model is described in Section \ref{sec2}. Section \ref{sec3} describes the typical MGI scenario reproduced in our simulations. Key results on the influence of injected impurity level are discussed in Section \ref{sec4}. Finally, a brief summary of the main results is presented in Section \ref{sec5}.

\section{Simulation model} 
\label{sec2}

Nonlinear simulation of EAST disruption induced by the He gas injection are carried out with the extended MHD NIMROD code, which integrates radiation and atomic physics model obtained from the KPRAD code~\cite{whyte1996}. This coupled model has been used to investigate rapid shutdown experiments in DIII-D~\cite{izzo2008, izzo2013, izzo2015}. The equations for the coupled model are as follows:
\begin{eqnarray}
\frac{\partial \vec{B}}{\partial t} = - \nabla \times \vec{E}
\label{eq:1}
\end{eqnarray}

\begin{eqnarray}
\vec{E}+\vec{V} \times \vec{B} = \eta \vec{J}
\label{eq:2}
\end{eqnarray}

\begin{eqnarray}
\rho \frac{d\vec{V}}{dt} = -\nabla p + \vec{J} \times \vec{B} + \nabla \cdot \mu \rho \nabla \vec{V}
\label{eq:3}
\end{eqnarray}

\begin{eqnarray}
n_{e} \frac{\partial T_{e}}{\partial t} = \left(\gamma -1 \right) \left[n_{e} T_{e} \nabla \cdot \vec{V} + \nabla \cdot \vec{q_{e}} - Q_{loss} \right]
\label{eq:4}
\end{eqnarray}

\begin{eqnarray}
\vec{q_{e}} = -n \left[\chi_{\parallel} \hat{b}\hat{b} +\chi_{\perp} \left(I-\hat{b}\hat{b} \right) \right] \cdot \nabla T_{e}
\label{eq:5}
\end{eqnarray}

\begin{eqnarray}
\frac{dn_{e}}{dt} + n_{e} \nabla \cdot \vec{V} = \nabla \cdot \left
(D\nabla n_{e} \right) + S_{ion/rec}
\label{eq:6}
\end{eqnarray}

\begin{eqnarray}
\frac{dn_{i}}{dt} + n_{i} \nabla \cdot \vec{V} = \nabla \cdot \left
(D\nabla n_{i} \right) + S_{ion/3-body}
\label{eq:7}
\end{eqnarray}

\begin{eqnarray}
\frac{dn_{Z}}{dt} + n_{Z} \nabla \cdot \vec{V} = \nabla \cdot \left
(D\nabla n_{Z} \right) + S_{ion/rec}
\label{eq:8}
\end{eqnarray}

Where, $\vec{E}$, $\vec{B}$, $p$, $\rho$, $\vec{V}$, $\vec{J}$, $n_{e}$, $n_{i}$, $n_{Z}$, $\eta$, $D$, $\mu$, $\chi_{\perp}$, $\chi_{\parallel}$, $S_{ion/rec}$, $S_{ion/3-body}$, $\gamma$, $Q_{loss}$, $\vec{q_{e}}$, and $T_{e}$ are the electric field, magnetic field, pressure, mass density, velocity, current density, number density of electrons, number density of ions, impurity number density, plasma resistivity, diffusivity, kinematic viscosity, heat flux coefficients in directions perpendicular and parallel to the magnetic field, density source from ionization and recombination, density source includes contribution from 3-body recombination, adiabatic index, energy loss, heat flux, and electron temperature, respectively. 
The equations (\ref{eq:2}) and (\ref{eq:3}) are similar to the usual NIMROD Ohm's law and momentum equation, except that the contributions from impurity species are included in the pressure and mass density. The resistivity is evaluated using Spitzer model, i.e. $\eta \propto T^{-3/2}$. Here, the quasi-neutrality condition is $n_{e}= n_{i}+\sum Zn_{z}$ in which $Z$ is the effective charge of the impurity ions.

The KPRAD model is used to update the charge-state populations of impurity species at each time step and at every grid point to evaluate source terms used in NIMROD temperature and density evolution equations. The total energy loss due to impurity includes contributions from bremsstrahlung, line radiation, ionization energy, and dilution induced isobaric plasma cooling. The ohmic heating is also present in the last term of equation (\ref{eq:4}). The injected impurity gas is supposed to be not completely ionized, as the three-body recombination can become important at low temperatures often encountered during MGI. The source/sink term of electrons contains contributions from both ionization and recombination of the impurities. The advection and diffusion terms are computed as a part of the NIMROD time advance.


\section{Typical MGI process from simulation}
\label{sec3}

\begin{figure}[htbp] 
\centering
\begin{minipage}{0.5\textwidth}
\includegraphics[width=1.0\textwidth]{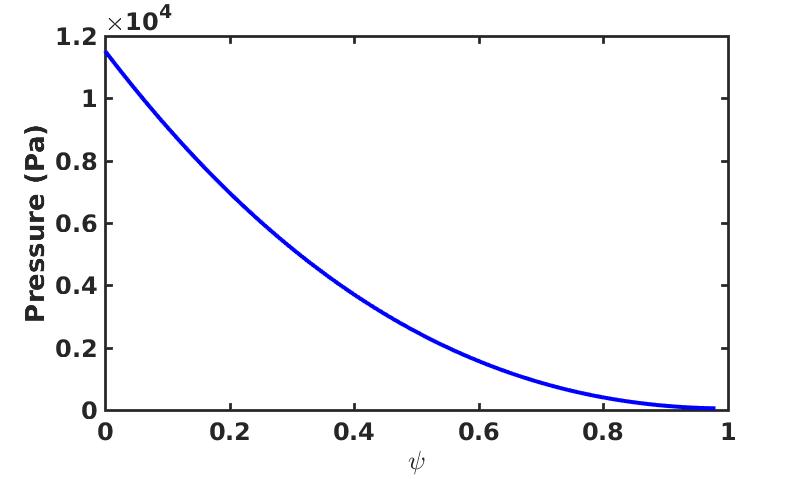}
\put(-190,110){\textbf{(a)}}
\end{minipage}
\begin{minipage}{0.5\textwidth}
\includegraphics[width=1.0\textwidth]{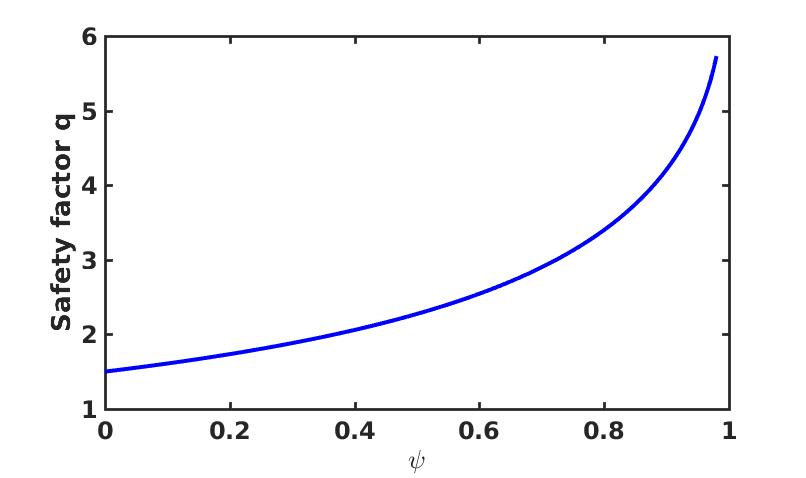}
\put(-190,110){\textbf{(b)}}
\end{minipage}
\caption{(a) Pressure and (b) safety factor profiles of the EAST equilibrium.} 
\label{fig_1} 
\end{figure}

\begin{figure}[htbp]
\centering
\includegraphics[width=0.7\textwidth]{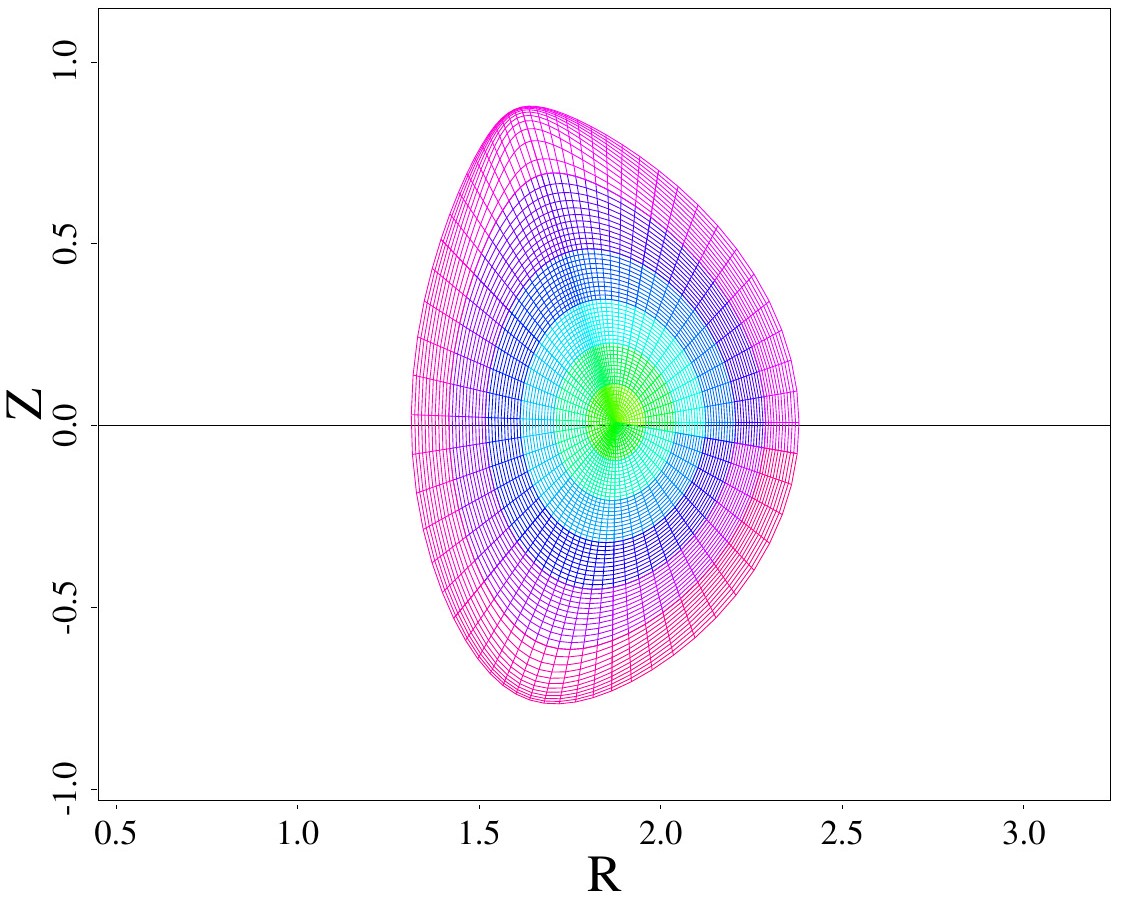}
\caption{2D finite element mesh for the EAST configuration used in the NIMROD simulations.}
\label{fig_2} 
\end{figure}

The equilibrium used in these simulations is constructed using the EFIT code based on the EAST discharge \# 71230 at the time of $4.8$ $s$. The pressure $(P)$ and safety factor $(q)$ equilibrium profiles are depicted in figure \ref{fig_1}. In polodial domain, we used $66 \times 66$ finite elements (figure \ref{fig_2}) with a polynomial degree of $5$. Other key equilibrium parameters used in the MGI simulations are listed in table \ref{table-1}.


\begin{table}
\begin{center}
\caption{\label{table-1} Main input parameters of MGI simulation}
\footnotesize
\begin{tabular}{@{}lll}
\br

Parameters & value & unit \\
\mr

 Resistivity ($\eta$)   & $2.8 \times 10^{-8}$     & $\Omega$ m  \\
 
 Lundquist number ($S$)   & $2.33 \times 10^{7}$     & - \\
 
 perpendicular thermal diffusion ($\chi_{\perp}$)   & $1$     & $m^{2}/s$  \\
 
 parallel thermal diffusion ($\chi_{\parallel}$)   & $1 \times 10^{10}$     & $m^{2}/s$  \\
 
 kinetic viscosity ($\mu$)   & $10$     & $m^{2}/s$  \\
 
 Particle diffusivity transport coefficient ($D$)   & $10$     & $m^{2}/s$  \\

 Core electron temperature ($T_{e}$)   & $1$     & $keV$  \\
 
 Core electron density ($n_{e}$)   & $3.5 \times 10^{19}$     & $m^{-3}$  \\
 
 Plasma current ($I_{P}$)   & $380$     & $kA$  \\
 
 Toroidal magnetic field ($B_{t}$)   & $1.8$     & $T$  \\
 
 Major radius ($R$)   & $1.85$     & $m$  \\
 
 Minor radius ($a$)   & $0.45$     & $m$  \\


\br
\end{tabular}\\
\end{center}
\end{table}
\normalsize

\begin{figure}[htbp]
\centering
\includegraphics[width=0.7\textwidth]{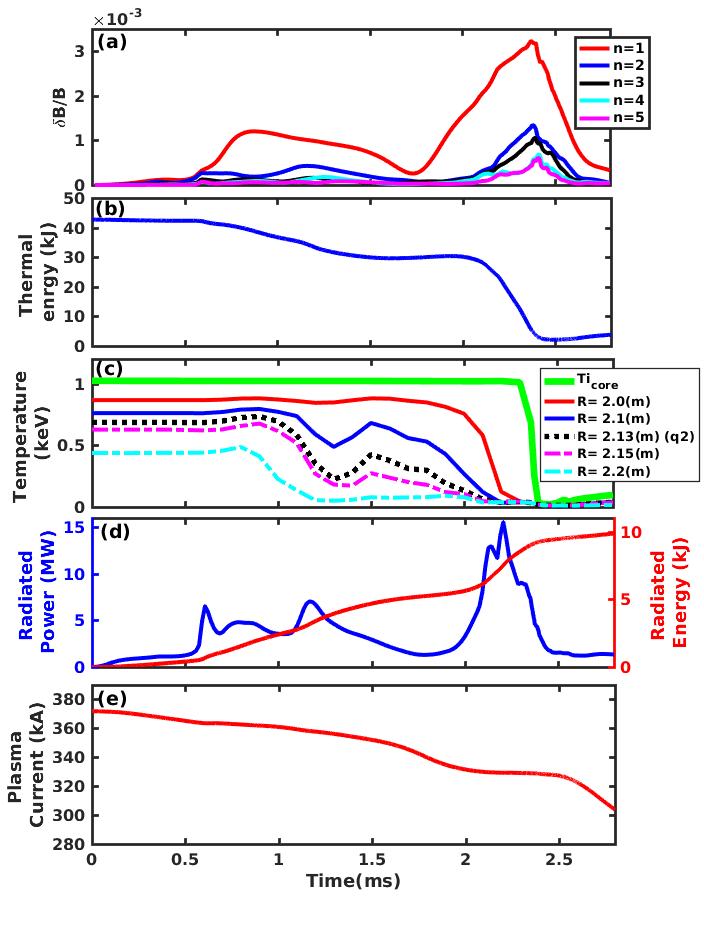}
\caption{Time evolution of major parameters in an MGI simulation: (a) magnetic amplitudes of $n = 1,2,3,4$ and $5$ modes (calculated as $\sqrt {\left[W_{mag,n}/W_{mag,0} \right]}$); (b) thermal energy; (c) temperature at different radial locations; (d) radiation power (left) and energy (right); and (e) plasma current.}
\label{fig_3} 
\end{figure}

Time evolution of some key quantities during the MGI triggered disruption show that the pre-TQ stage lasts for $2.3$ $ms$ which ends when the plasma core temperature starts to decrease (figure \ref{fig_3}). Initially, the injected impurities are mostly distributed at the edge and then slowly diffuse towards the core. The radiated power ($P_{rad}$) is therefore initially less intense due to low boundary temperature. As the cold front advances inwards and the impurities penetrates deeper, the magnitude of $P_{rad}$ increases significantly. A conspicuous peak in $P_{rad}$ is initially observed at $0.6$ $ms$ (figure \ref{fig_3}(d)), followed by the first MHD activity (figure \ref{fig_3}(a)), which saturates around $0.85$ $ms$ and coincides with the decrease of plasma temperature outside of $q = 2$ surface (i.e. the temperature drops at the radial locations $2.2$ $m$ and $2.15$ $m$) as shown in the figure \ref{fig_3}(c). After a small delay, another spike in the radiated power occurs ($t \sim 0.95$ ms), presumably due to the cold front penetration within the $q = 2$ surface. As soon as the first MHD burst (with dominant $n=1$ mode) tends to decrease, the temperature collapse at $q = 2$ and outside is initiated.


Subsequently, a second major MHD activity is triggered at $t \sim 1.8$ ms, where the magnetic perturbations of all modes grow substantially. Meanwhile, the $P_{rad}$ surges again to the maximum value of about $15$ MW (figure \ref{fig_3}(d)). The peak MHD activity and radiation power are immediately followed by the onset of TQ, when the core temperature starts to decrease at $2.3$ ms (figure \ref{fig_3}(c)). The magnitude of second MHD activity is significantly larger than the first one, leading to the complete collapse of core temperature (figure \ref{fig_3}(c)). Figure \ref{fig_3}(a) reveals that the majority of plasma thermal energy (more than $60 \%$) is lost during the pre-TQ phase due to cooling of the edge region. The radiated energy at the end of the TQ is $10 kJ$ ($\sim 25\%$ of lost thermal energy) which is comparable to the EAST experimental value. The TQ phase is followed by the current quench phase (CQ), when the plasma current decays due to increased plasma resistivity, and a slight current spike at $2.5$ $ms$ is identified right prior to the start of CQ (figure \ref{fig_3}(e)).


\begin{figure}[htbp]
\centering
\includegraphics[width=0.6\textwidth]{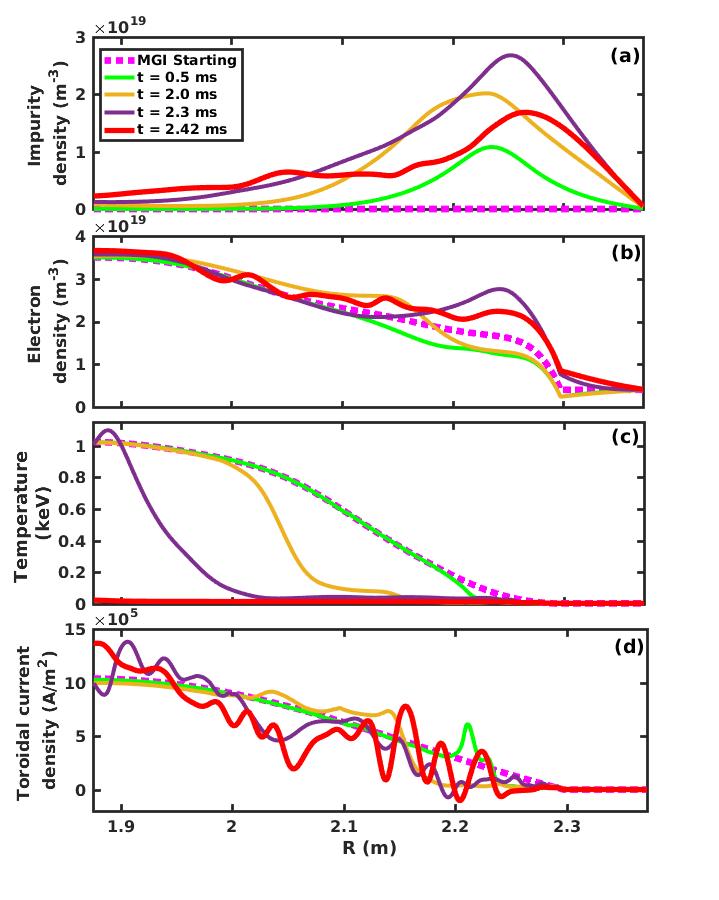}
\caption{Time evolution of radial profiles for: (a) impurity density; (b) electron density; (c) ion temperature; and (e) toroidal current density during pre-TQ and TQ phases.}
\label{fig_4} 
\end{figure}

\begin{figure}[htbp] 
\centering
\begin{minipage}{0.45\textwidth}
\includegraphics[width=1.0\textwidth]{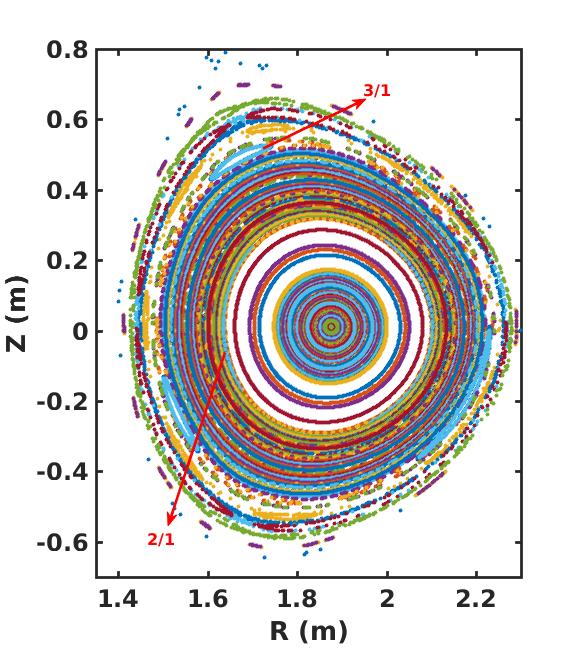}
\put(-163,195){\textbf{(a)}}
\end{minipage}
\begin{minipage}{0.45\textwidth}
\includegraphics[width=1.0\textwidth]{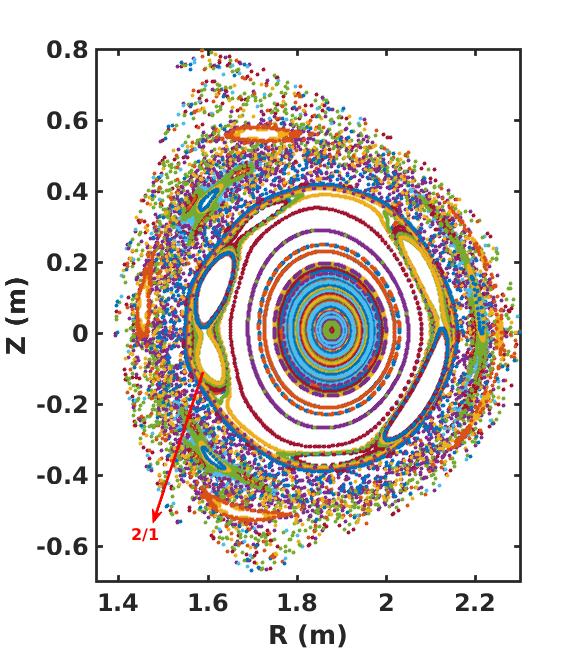}
\put(-163,195){\textbf{(b)}}
\end{minipage}
\begin{minipage}{0.45\textwidth}
\includegraphics[width=1.0\textwidth]{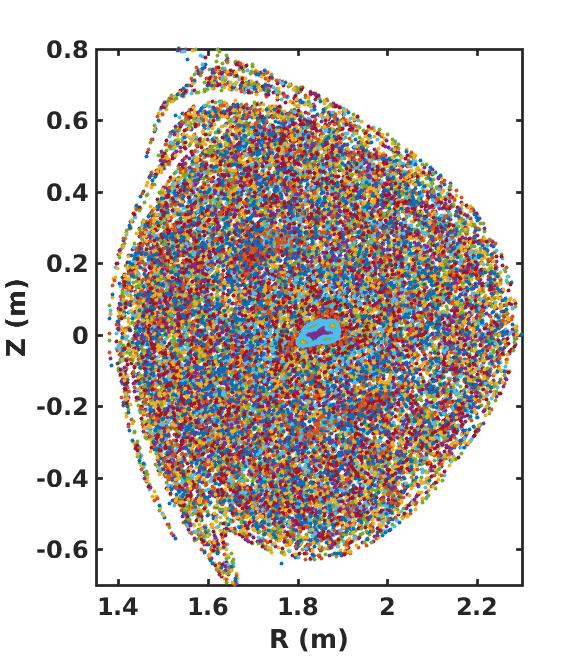}
\put(-163,195){\textbf{(c)}}
\end{minipage}\begin{minipage}{0.45\textwidth}
\includegraphics[width=1.0\textwidth]{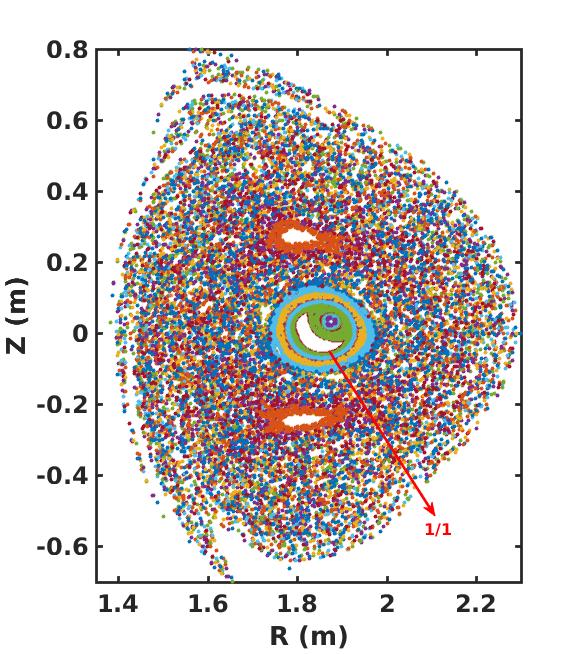}
\put(-163,195){\textbf{(d)}}
\end{minipage}
\caption{Poincaré plots in poloidal plane at: (a) $t = 0.5$ ms; (b) $t = 2.0$ ms; (c) $t = 2.42$ ms (the end of TQ); (d) $t = 2.5$ ms (the start of CQ).}
\label{fig_5}
\end{figure}

 The time evolution of plasma profiles (impurity density, electron density, temperature, toroidal current density) and magnetic flux surfaces (Poincaré plots) during different disruption stages is illustrated in figures \ref{fig_4} and \ref{fig_5}. The injected impurity enhances the electron density at the edge and cool down the plasma at the outer boundary (figures \ref{fig_4}(b) and (c)). The plasma resistivity increases throughout the cooled-down region, leading to contraction of current profile, which triggers MHD instabilities such as $m/n = 3/1$ and $2/1$ modes ($m$ and $n$ are the polodial and the toroidal mode numbers respectively) (figure \ref{fig_5}(a)). However, the $2/1$ island width at this point is not prominent. As more impurity density accumulates around the $q = 2$ surface, the current density channel shrinks inside $q = 2$ (figure \ref{fig_4}(d)) and contributes to a rapid growth of $2/1$ magnetic island (figure \ref{fig_5}(b)). By the time $t= 2.0$ ms, the magnetic flux near the $q = 2$ surface and its periphery outside become stochastic, whereas the magnetic surfaces inside down to the magnetic axis remain intact. At the end of the TQ, core flux surfaces are destroyed and almost the entire plasma domain becomes magnetically stochastic (figure \ref{fig_5}(c)). The $1/1$ magnetic island structure appears at the beginning of the CQ stage (figure \ref{fig_5}(d)). The CQ phase is not elaborated further here as it is beyond the scope of this study.
 
\begin{figure}[htbp]
\centering
\includegraphics[width=0.6\textwidth]{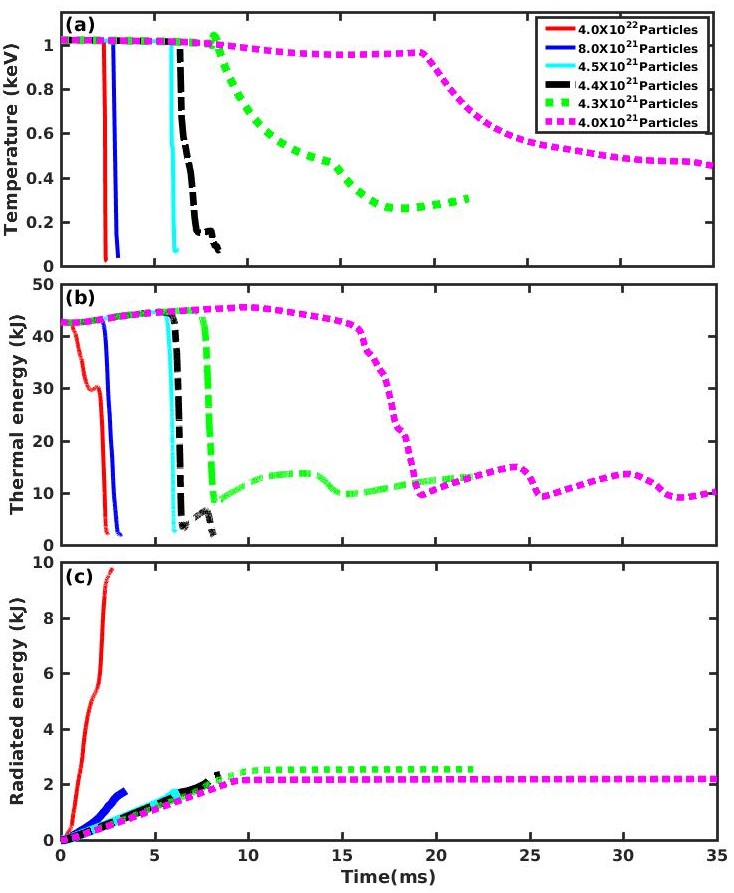}
\caption{Time evolution of: (a) core temperature drop; (b) thermal energy; and (c) radiation energy for different injection levels.}
\label{fig_6} 
\end{figure}



\section{Influence of impurity injection level}
\label{sec4}

Systematic simulations are performed for different injection levels to investigate its effect on thermal collapse. Plasma equilibrium parameters are kept same in all simulations (table \ref{table-1}), only the amount of injected impurities are varied. A critical injection level $(4.4 \times 10^{21} particles)$ is identified, around which the TQ behavior changes distinctively. As demonstrated in figure \ref{fig_6}(a) in the cases with injected $He$ above the critical injection level, the electron temperature at magnetic axis drops below several tens of $eVs$ in a single step (i.e. major disruption), whereas in the case of impurity injection below the critical level, the complete drop of core temperature takes several steps after multiple fractional decrements (i.e. minor disruption). Furthermore, it is observed that the duration of pre-TQ phase is largely affected by the amount of impurity injection. By reducing the amount of injected impurities, the precursor phase stays longer, i.e. more time is required for the MHD modes to saturate and induce TQ. The lost thermal energy in each case is around  90-95 $\%$ at the end of TQ, which is significantly more than the radiated energy. In the larger injection case ($N_{He}=4 \times 10^{22} particles$) the radiated energy is about $10 kJ$, while for the smaller injection cases ($N_{He}=4 \times 10^{21} particles$), it is approximately $2 kJ$ (figure \ref{fig_6}(b)). This implies that a relatively small fraction of thermal energy is dissipated away through radiation due to the low-Z charge of $He$ gas, which may be not an efficient radiator.

\begin{figure}[htbp] 
\centering
\begin{minipage}{0.50\textwidth}
\includegraphics[width=8cm, height=11cm]{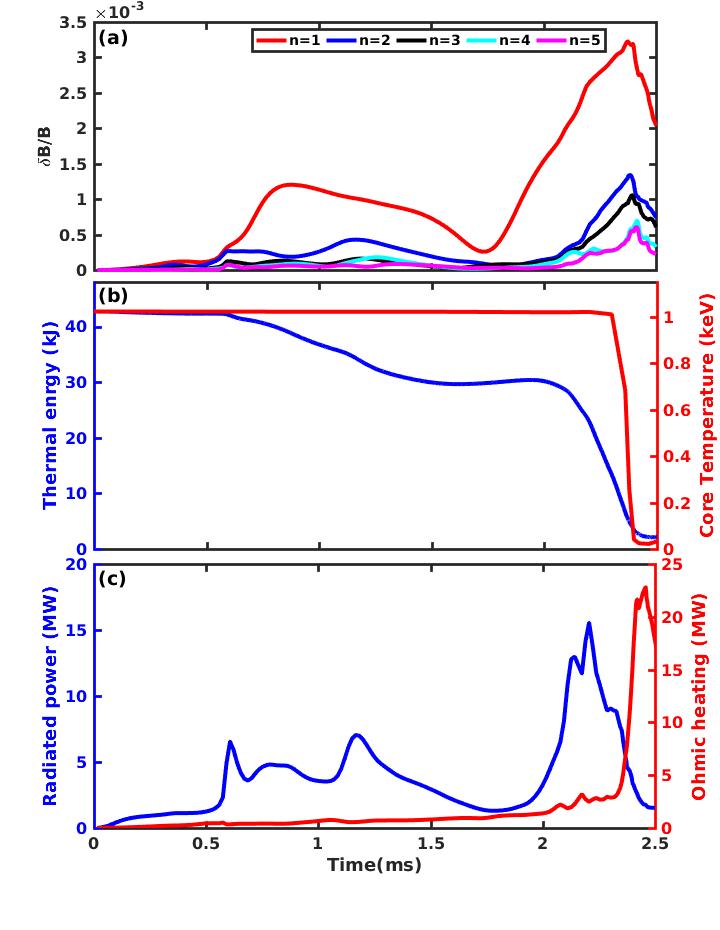}
\put(-225,305){\textbf{(I)}}
\end{minipage}
\begin{minipage}{0.47\textwidth}
\includegraphics[width=8cm, height=11cm] {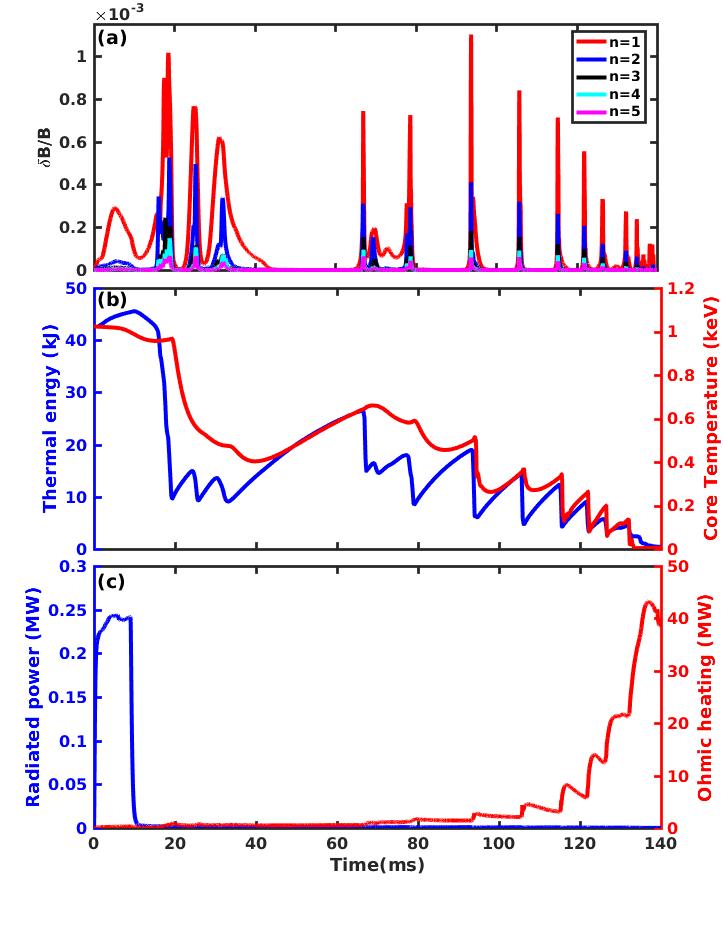}
\put(-230,300){\textbf{(II)}}
\end{minipage}
\caption{Time evolution of (a) normalized magnetic energy; (b) thermal energy (left) and core temperature (right); (c) radiated power (left) and Ohmic heating (right); for Helium injection levels above (I) and below (II) the critical value.}
\label{fig_7} 
\end{figure}


\newpage
\begin{figure}[htbp] 
\centering
\begin{minipage}{0.23\textwidth}
\includegraphics[width=1.0\textwidth]{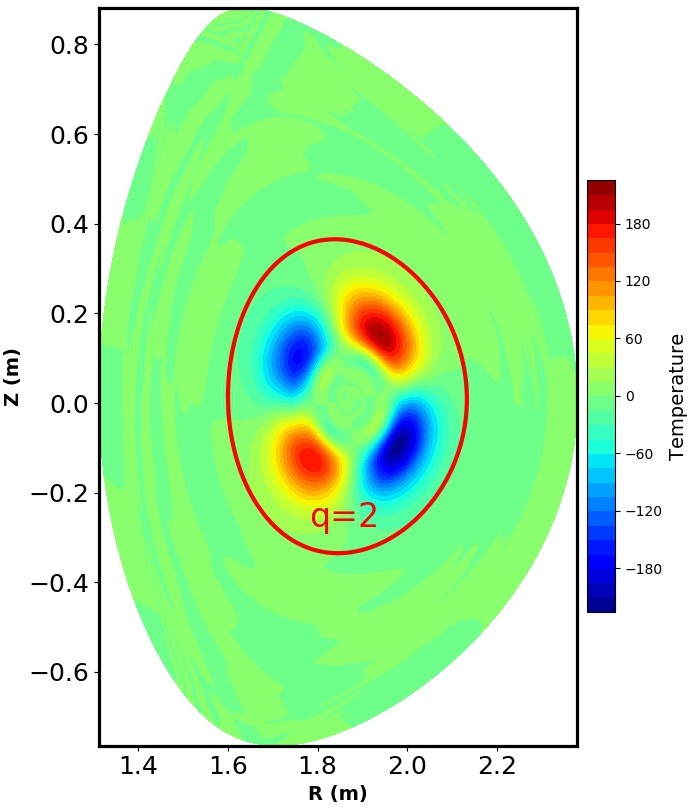}
\put(-75,105){\texttt{(a1)}}
\put(-65,15){\texttt{n1}}
\end{minipage}
\begin{minipage}{0.23\textwidth}
\includegraphics[width=1.0\textwidth]{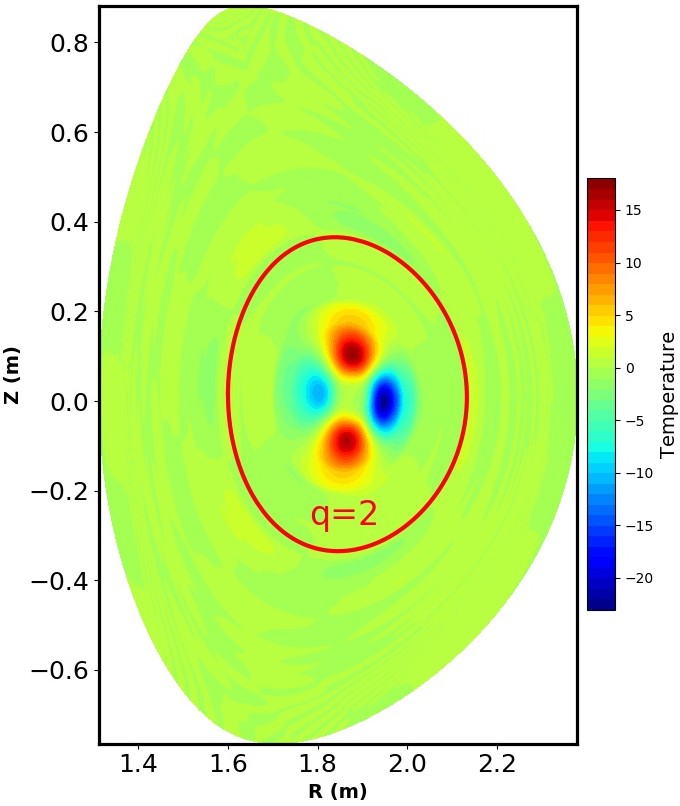}
\put(-75,105){\texttt{(b1)}}
\put(-65,15){\texttt{n1}}
\end{minipage}

\begin{minipage}{0.23\textwidth}
\includegraphics[width=1.0\textwidth]{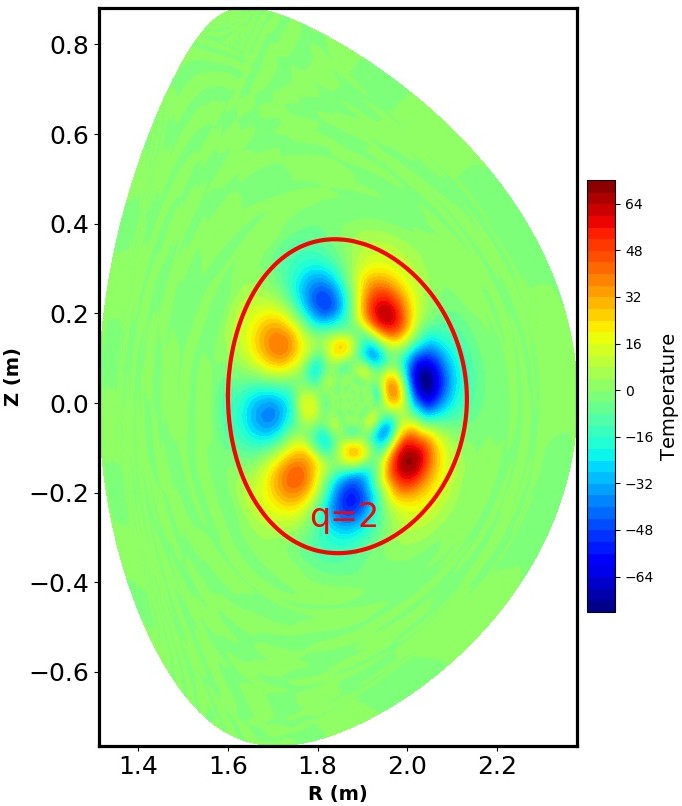}
\put(-75,105){\texttt{(a2)}}
\put(-65,15){\texttt{n2}}
\end{minipage}
\begin{minipage}{0.23\textwidth}
\includegraphics[width=1.0\textwidth]{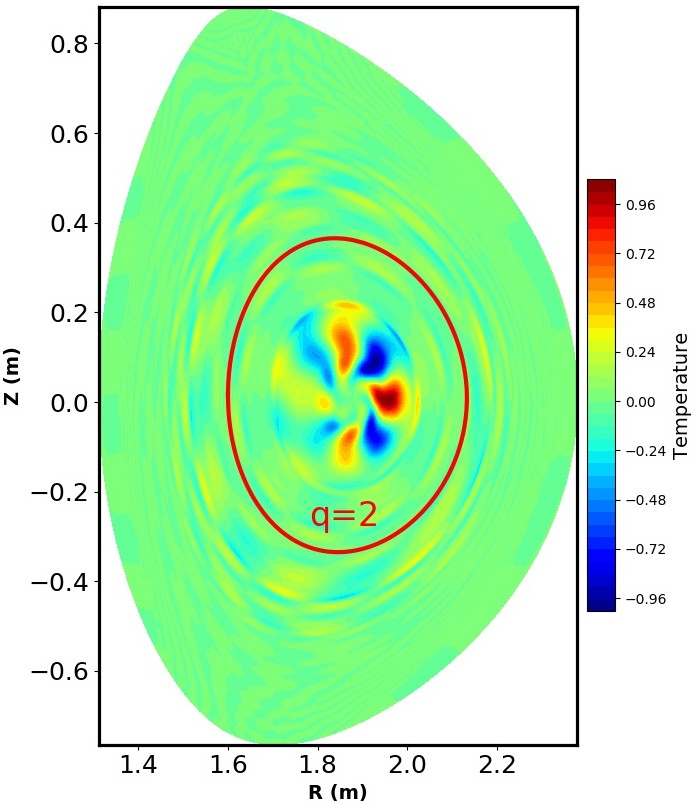}
\put(-75,105){\texttt{(b2)}}
\put(-65,15){\texttt{n2}}
\end{minipage}

\begin{minipage}{0.23\textwidth}
\includegraphics[width=1.0\textwidth]{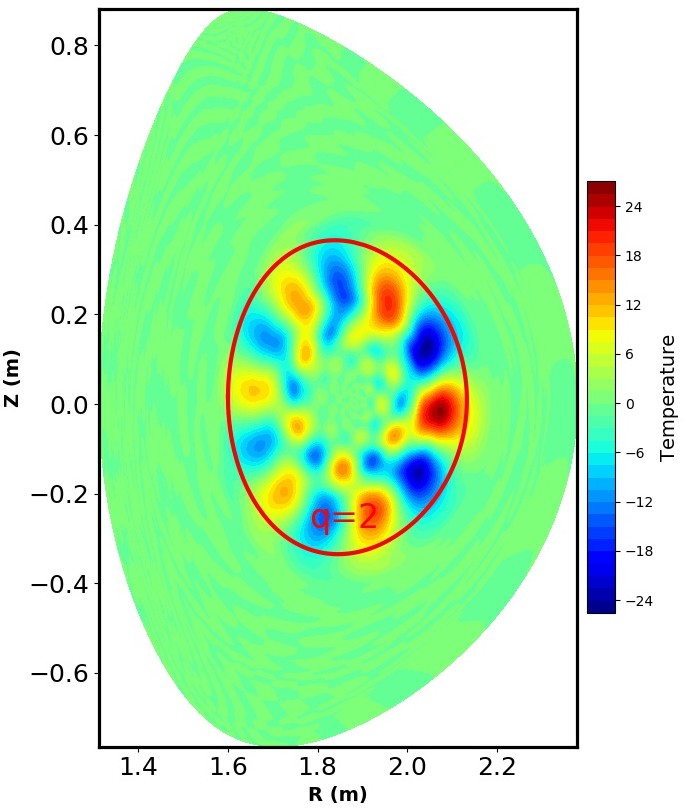}
\put(-75,105){\texttt{(a3)}}
\put(-65,15){\texttt{n3}}
\end{minipage}
\begin{minipage}{0.23\textwidth}
\includegraphics[width=1.0\textwidth]{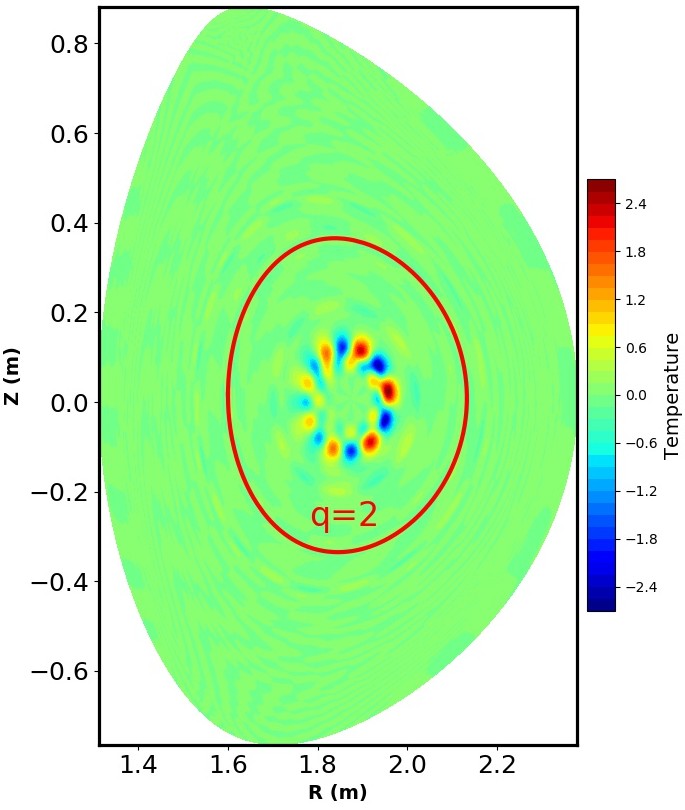}
\put(-75,105){\texttt{(b3)}}
\put(-65,15){\texttt{n3}}
\end{minipage}

\begin{minipage}{0.23\textwidth}
\includegraphics[width=1.0\textwidth]{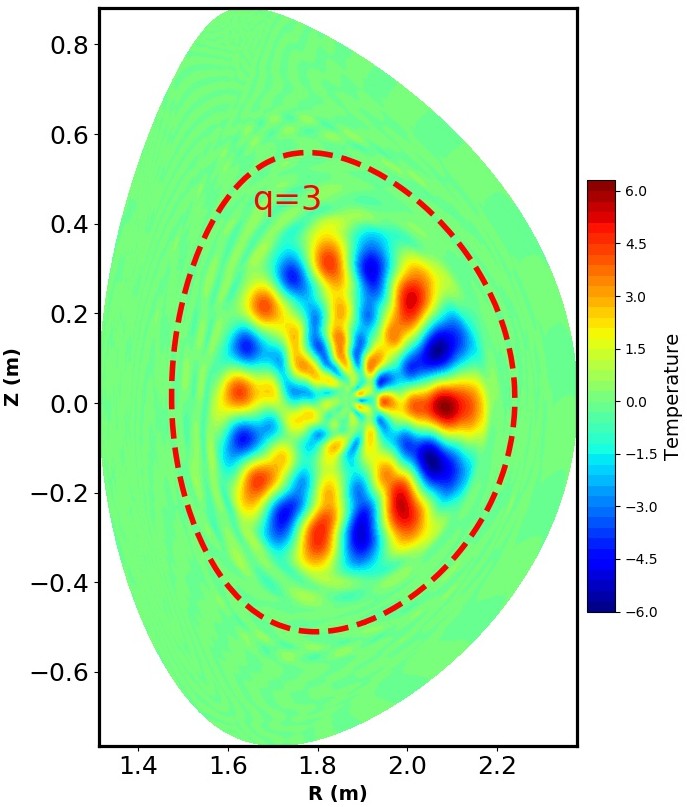}
\put(-75,105){\texttt{(a4)}}
\put(-65,15){\texttt{n4}}
\end{minipage}
\begin{minipage}{0.23\textwidth}
\includegraphics[width=1.0\textwidth]{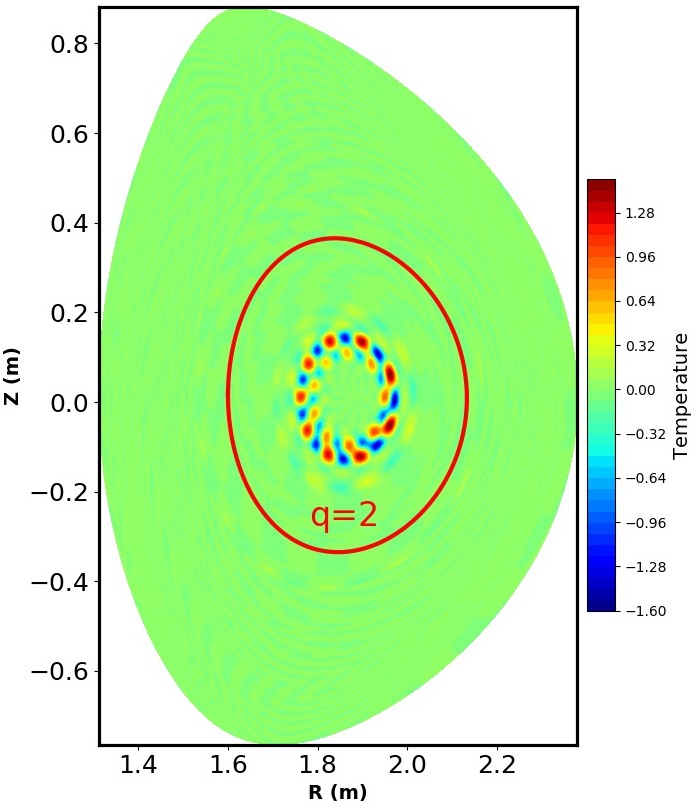}
\put(-75,105){\texttt{(b4)}}
\put(-65,15){\texttt{n4}}
\end{minipage}

\begin{minipage}{0.23\textwidth}
\includegraphics[width=1.0\textwidth]{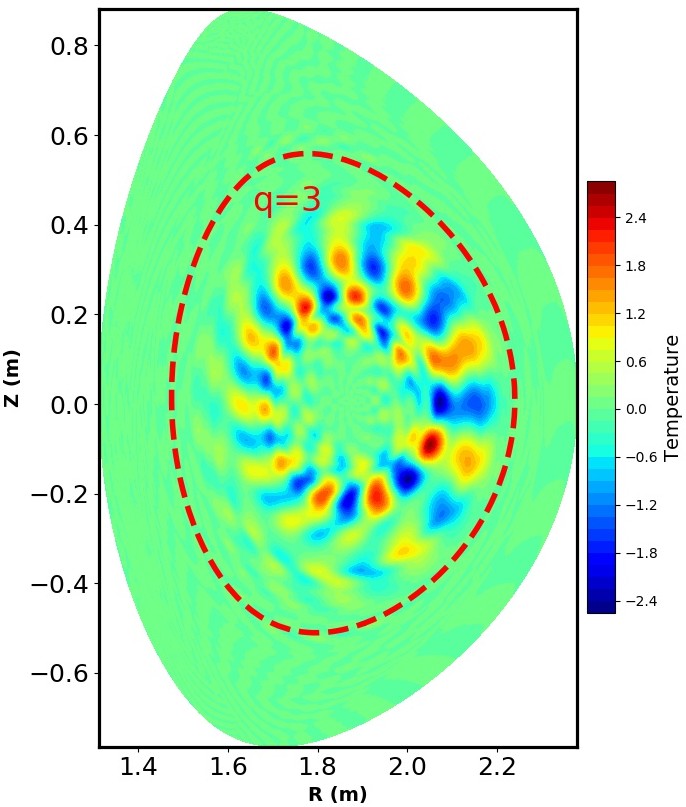}
\put(-75,105){\texttt{(a5)}}
\put(-65,15){\texttt{n5}}
\end{minipage}
\begin{minipage}{0.23\textwidth}
\includegraphics[width=1.0\textwidth]{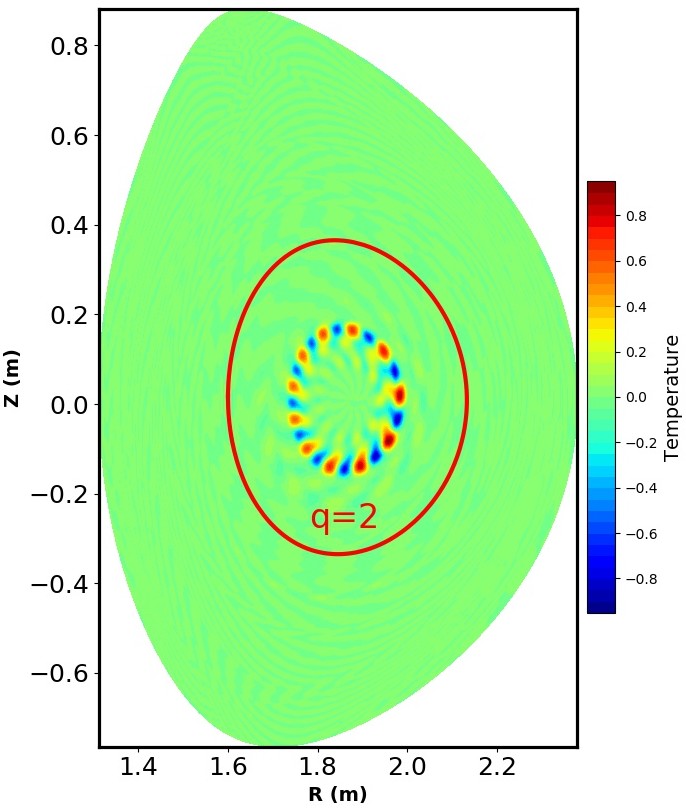}
\put(-75,105){\texttt{(b5)}}
\put(-65,15){\texttt{n5}}
\end{minipage}
\caption{Temperature contours of all toroidal modes with the toroidal mode numbers $n = 1-5$ for Helium injection level above ((a1)-(a5), t = 1.3 ms) and below ((b1)-(b5), t=  19.5 ms) the critical value.}
\label{fig_8}
\end{figure}
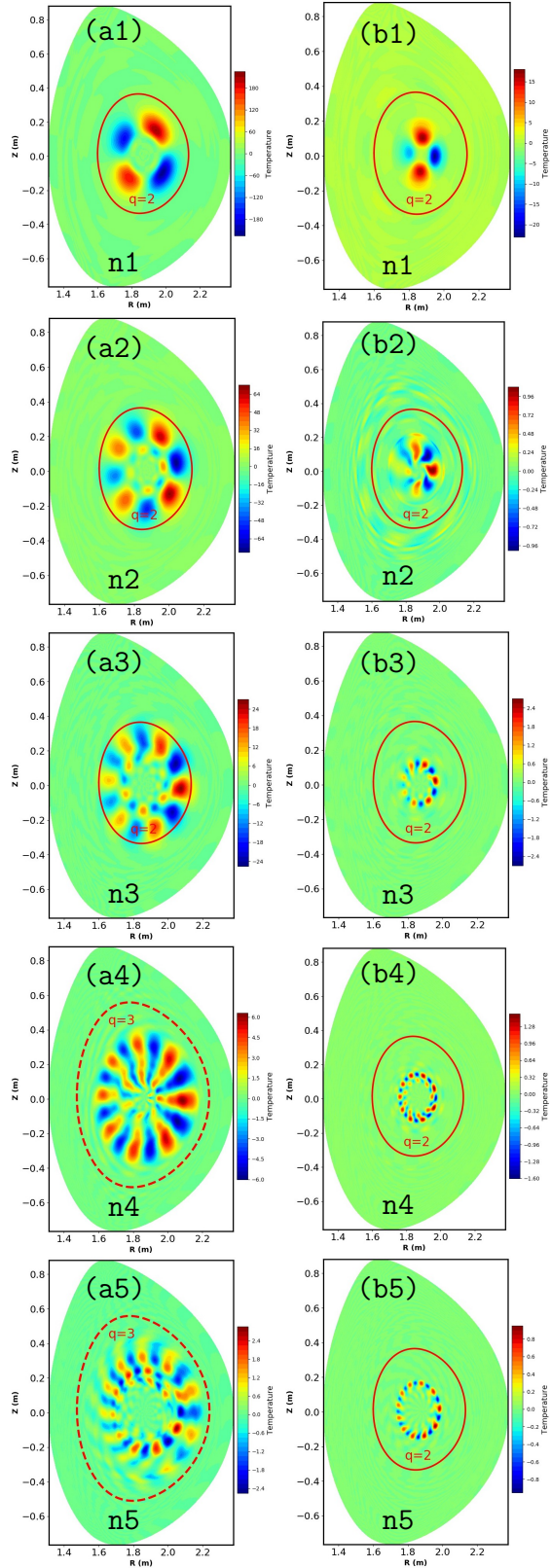



The degree of core temperature collapse is another key to distinguishing between major and minor disruptions. A disruption is triggered when the magnetic perturbation grows to sufficiently high amplitude, which is influenced by the amount of injected impurities. The magnetic perturbation level in turn affects the rate of perpendicular heat transport and the speed of drop in core temperature. To further elucidate the influence of injected impurity levels, time evolution of key plasma parameters like normalized magnetic energy, thermal energy, core temperature, radiated power and ohmic heating power are compared side-by-side for the higher (i.e. major disruption) and the lower impurity injection (i.e. minor disruption) cases (figure \ref{fig_7}).
Significant differences in the amplitude of magnetic perturbation and radiated power are observed. Due to the stronger radiation power in the higher injection case, the TQ phase is much shorter in duration ($\sim$ 0.25 $ms$) and takes place in one single step (figure  \ref{fig_7}(I)). In contrast, in the sub-critical lower injection case, the collapse of core temperature and stored thermal energy takes much longer time ($\sim$ 100 $ms$) to complete after multiple cycles of minor collapse and recovery (figure \ref{fig_7}(II)). In both cases, the thermal collapses correlate closely with the spikes of magnetic perturbations in both timing and magnitude, suggesting the roles of MHD activities in the effects of injection level on the TQ process. The contributions of MHD activity can be further revealed from the comparison of the mode structure in the temperature perturbation contours in the poloidal planes for the two representative injection cases (figure \ref{fig_8}). Whereas in the higher injection level case, the MHD mode structures are radially extended between the $q=1$ and the $q=2$ surfaces, the mode structures are mostly localized around the $q=1$ surface in the lower injection level case.


\section{Summary}
\label{sec5}

In summary, a critical injection level has been identified in the simulations of the EAST MGI process. It has been demonstrated that a strong MHD burst leads to the destruction of global confinement and achieve complete thermal collapse in a single step (major disruption) when injection level is above the critical level. However, in case of injected $He$ gas amount below the critical level, the cooling from impurity radiation is relatively weak, resulting in smaller amplitude of MHD activity. As a result, there are several cycles of partial collapses (i.e. minor disruptions) and recoveries until the core temperature drops to a few tens of $eVs$ (i.e. complete TQ). For both high and low injection levels, the magnitude, the mode structure, and the timing of MHD activities correlate closely with those of the thermal collapse events during the MGI process. Such a correlation signifies the role of the impurity radiation induced MHD instability as the key underlying mechanism that governs the disruption mitigation physics.

The development of a predicate model or scaling for the critical injection level would be helpful for the design and practice of MGI disruption mitigation scheme, which may require more quantitative understanding of the impurity injection and the induced MHD processes. We plan on looking into this direction in future studies.

\ack

This research was supported by the National Magnetic Confinement Fusion Science Program of China Grant NO.2019YEE03050004, the National Natural Science Foundation of China Grant Nos.11775221 and 51821005, U.S.DOE Grant Nos.DE-FG02-86ER53218 and DESC0018001, and the Fundamental Research Funds for the Central Universities at Huazhong University of Science and Technology Grant No.2019kfyXJJS193. We are grateful for the support from NIMROD team, and to Drs. Songtao Mao and Yanmin Duan of the EAST experimental group for providing the initial equilibrium data. This research used the computing resources from the Supercomputing Center of University of Science and Technology of China. The author Abdullah Zafar acknowledges the support from the Chinese Government Scholarship.

\newpage

\section*{References}
\providecommand{\newblock}{}

\end{document}